\documentclass{aa}

\usepackage{graphicx}
\usepackage{natbib}
\usepackage{txfonts}

\bibpunct{(}{)}{;}{a}{}{,}

\begin{document}

\title{Experimental \ion{Mg}{i} oscillator strengths and radiative lifetimes for astrophysical applications on metal-poor stars}
\subtitle{New data for the \ion{Mg}{i} b triplet}

\author{M. Aldenius\inst{1}
\and J.D. Tanner\inst{2}
\and S. Johansson\inst{1} 
\and H. Lundberg\inst{3}
\and S.G. Ryan\inst{2,4}}

\institute{Atomic Astrophysics, Lund Observatory, Lund University, Box 43, SE-221 00 Lund, Sweden\\
\email{maria@astro.lu.se}
\and Department of Physics and Astronomy, The Open University, Walton Hall, Milton Keynes, MK7 6AA, United Kingdom
\and Atomic Physics, Department of Physics, Lund Institute of Technology, Box 118, SE-221 00 Lund, Sweden
\and Centre for Astrophysics Research, STRI, University of Hertfordshire, College Lane, Hatfield, AL10 9AB, United Kingdom}

\date{Received 18 August 2006 / Accepted 26 September 2006}

\authorrunning{Aldenius et al.}
\titlerunning{Experimental \ion{Mg}{i} oscillator strengths}


\abstract
{The stellar abundance ratio of Mg/Fe is an important tool in diagnostics of galaxy evolution. In order to make reliable measurements of the Mg abundance of stars, it is necessary to have accurate values for the oscillator strength ($f$-value) of each of the observable transitions. In metal-poor stars the \ion{Mg}{i} 3p-4s triplet around 5175~\AA\ (Fraunhofer's so-called b lines) are the most prominent magnesium lines. The lines also appear as strong features in the solar spectrum.}
{We present new and improved experimental oscillator strengths for the optical \ion{Mg}{i} 3p-4s triplet, along with experimental radiative lifetimes for six terms in \ion{Mg}{i}. With these data we discuss the implications on previous and future abundance analyses of metal-poor stars.}
{The oscillator strengths have been determined by combining radiative lifetimes with branching fractions, where the radiative lifetimes are measured using the laser induced fluorescence technique and the branching fractions are determined using intensity calibrated Fourier Transform (FT) spectra. The FT spectra are also used for determining new accurate laboratory wavelengths for the 3p-4s transitions.}
{The $f$-values of the \ion{Mg}{i} 3p-4s lines have been determined with an absolute uncertainty of 9\,\%, giving an uncertainty of $\pm 0.04$\,dex in the log\,$gf$ values. Compared to values previously used in abundance analyses of metal-poor stars, rescaling to the new values implies an increase of typically $0.04$\,dex in the magnesium abundance.}
{}


\maketitle



\section{Introduction}

The key to understanding the evolution of galaxies including our own Galaxy, the Milky Way, is through measurements of the abundances of the chemical elements. The chemical compositions of galaxies change as different generations of stars form, evolve and later burn out. At the end of their lives, sufficiently massive stars eject newly synthesised elements into the gaseous interstellar medium of their host galaxy. New stars form from this gas reservoir with a composition enriched in heavy elements compared to previous stellar generations. By measuring the diverse compositions of the stars in a galaxy, it is possible to infer its evolutionary history.

Amongst the elements whose abundances can be measured, two are particularly important: magnesium and iron. Iron is the final product of exothermic nuclear burning, and is produced in two sites: the cores of massive stars preceding their collapse into supernovae of type SN Ib, Ic and II, and in the fireballs that become the low-mass supernovae of type SN Ia. Magnesium is also produced by massive stars that later become core-collapse supernovae, but it is produced much earlier in the star's evolution and under much more sedate conditions, in the non-explosive burning of neon. Massive stars are the sole source of Mg, whereas Fe is produced in the supernovae of high- and low-mass stars, though predominantly the latter.

From observational and theoretical studies of stars in the Milky Way, astronomers have discovered that the abundance ratio of Mg/Fe is an important diagnostic of galaxy evolution. The ratio reduces during later stages of galaxy evolution due to the late onset of Fe production in SN Ia. Hence observations of Mg and Fe together constrain the lifetimes of the stellar populations that make up galaxies and their star-formation histories. This is seen both in our own Galaxy, where star-by-star measurements of the Mg/Fe ratio is possible \citep{Gratton00}, and in more distant elliptical galaxies \citep{Matteucci98} where the combined light of many stars shows a higher value of the Mg/Fe ratio than in, for comparison, the Sun.

The Mg/Fe ratio also provides very important information on the mechanism of the SN II explosion. Although this certainly involves the collapse of the core of a massive star, astronomers have been unable to calculate where in the core the imploding and the exploding layers divide - the location of the so-called mass cut. This in turn leaves uncertainty over the fate of the compact remnant, i.e. whether it becomes a neutron star or a black hole, and the amount of newly synthesised material that escapes to enrich the interstellar medium and future generations of stars. The Mg/Fe ratio can shed light on this situation. Mg is produced well outside the core of the progenitor of the SN II, whereas Fe is produced in the region of the mass cut. Observations of the Mg/Fe abundance ratio in stars that have formed from gas ejected by SN II therefore show how much of the Fe core has been ejected compared to the amount of Mg produced earlier in its evolution. Variations in this ratio from one SN II to another, for example as a result of different progenitor masses, can also be estimated from an observational analysis of the range of Mg/Fe ratios preserved in later generations of stars. For reasons associated with the uncertainty over Fe production, many astronomers would prefer to use Mg as the usual standard for measuring the overall heavy-element content of a star, whereas Fe is used currently because of the greater ease of measuring (but not interpreting) that atomic species.

In order to make reliable measurements of the Mg abundances of stars, it is necessary to have accurate values for the oscillator strength (\emph{f}-value) of each of the observable transitions. As stars vary widely in temperature and Mg content, the strengths of the Mg spectral lines also vary greatly such that lines that are easily measured and analysed in one star may be unsuitable for the task in another. For this reason, it is vital that accurate oscillator strengths are available for a wide range of observable transitions. In the optical region where most stellar spectra are obtained, the oscillator strength has, to our knowledge, only been measured in the laboratory for one \ion{Mg}{i} line, the  3s$^2\,^1$S-3s3p\,$^3$P$_1$ intercombination transition at 4571 \AA\ \citep{Kwong82}. This is the only ground state transition of magnesium in the optical region, and due to the strong LS coupling in \ion{Mg}{i} it is very weak, log~\emph{gf}=-5.69. Measurements of oscillator strengths for other \ion{Mg}{i} transitions have been reported by \citet{Ueda82}, but these were not purely experimental, since the absolute scale of the \emph{f}-values was taken from theoretical calculations \citep{Wiese69}. Thus, no purely experimental \emph{f}-values exist for transitions between excited levels, even though there are a number of laser measurements of radiative lifetimes particularly for singlet states \citep[see e.g.][]{Jonsson84,Larsson93}. The reason is that these levels have branches at infrared wavelengths, which requires experimental branching fractions of IR and optical lines to get absolute \emph{A}-values and \emph{f}-values.

\begin{figure}
 \includegraphics[width=84mm]{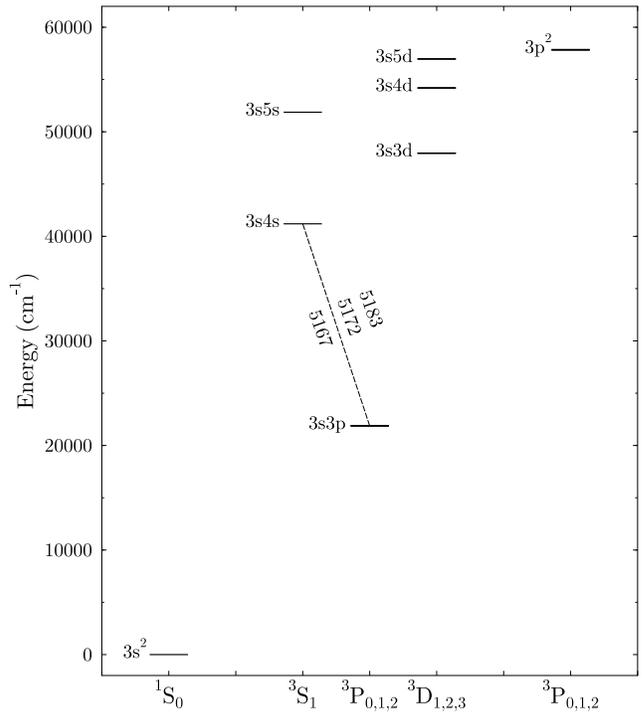}
 \caption{Partial energy level diagram of \ion{Mg}{i} displaying the ground state, the 3s3p\,$^3$P levels and the levels for which the lifetimes have been measured in this work. The 3p-4s lines are marked with wavelengths in \AA .}
 \label{diagramfig}
\end{figure}

In this paper we present new laboratory measurements of radiative lifetimes of the 3s4s\,$^3$S, 3s5s\,$^3$S, 3s3d\,$^3$D, 3s4d\,$^3$D, 3s5d\,$^3$D and 3p$^2$\,$^3$P terms in \ion{Mg}{i} using the laser induced fluorescence technique. A partial energy level diagram of the involved levels is displayed in Fig.~\ref{diagramfig}. One of these LS terms, 3s4s\,$^3$S, has only one decay route, down to 3s3p\,$^3$P, and the corresponding transitions appear in the optical region. By measuring the branching fractions, using Fourier Transform (FT) spectrometry, for these transitions we have derived absolute oscillator strengths for three 3p-4s lines (multiplet number 2) around 5175~\AA. These three lines, Fraunhofer's so called b lines, appear as strong features in the solar spectrum as well as appearing in spectra of metal-poor stars and can be used for abundance determinations. The implications of the new data for abundance analyses of metal-poor stars are discussed in Sect.~\ref{astrophyssec}.

\section{Radiative lifetimes}

The radiative lifetimes were measured with the method of selective excitation of an atomic level by wavelength-tuned pulsed laser radiation and time-resolved detection of radiation released at the subsequent decay. Free magnesium atoms were produced by focusing a pulsed Nd:YAG laser beam onto a rotating Mg target. The laser pulse created a small plasma, expanding from the target and containing electrons, atoms and ions of various ionisation stages. The laser pulse had duration of 10 ns and typical energy of 10 mJ. This atomic source has the advantage of high particle density and the possibility of using populated metastable levels as starting points for laser excitation. Measurements were performed on atoms in the later and slower part of the plasma by adjusting the time delay of the excitation pulse.

The investigated levels have energies in the range 40000-60000~cm$^{-1}$ and in the measurements the metastable 3s3p\,$^3$P term was utilised as the starting point for the selective excitation. Pulsed tuneable laser light was produced by a Nd:YAG laser system. The 10 ns long pulses from an injection-seeded frequency-doubled Nd:YAG laser were first compressed to 1 ns using stimulated Brillouin scattering in a water cell. The compressed pulse pumped a dye laser operated with red dyes. The UV wavelengths for excitation were then produced by frequency-doubling in a non-linear crystal and Raman wavelength-shifting in a hydrogen cell.

The excitation beam interacted with the magnesium atoms about 1~cm above the target. Fluorescent light released at the decay was captured using a 0.25 m vacuum monochromator and a multichannel-plate photo-multiplier with a rise time of 0.2 ns. The photo-multiplier was connected to a digital transient recorder with an analogue bandwidth of 1~GHz and real-time sampling rate of 2~GSamples/s. The temporal structure of the VUV excitation pulses was recorded with the same detection system by inserting a metal rod into the place of the plasma. In the measurements laser pulses and fluorescence signals were recorded alternately. The lifetimes were evaluated by fitting the fluorescence signal to a convolution of the laser pulse and an exponential. The procedure takes the limited time response into account since the recorded pulse and fluorescence are affected by the same response function. The final lifetime values are averages from a number of such recordings. The error bars of the values include the equal parts of statistical scattering between different recordings and the possibility of disregarded systematic effects. The lifetime results are given in Table~\ref{Lifetable}, where the number in brackets gives the uncertainty in the final digit(s). The fine- structure in the $^3$D states could not be resolved in the measurements. For further discussion on the technique of measuring radiative lifetimes with laser induced fluorescence see e.g. \citet{Li00}.

\begin{table}
\begin{minipage}[t]{\columnwidth}
\caption{Experimental radiative lifetimes of \ion{Mg}{i}}
\label{Lifetable}
\centering
\renewcommand{\footnoterule}{}
\begin{tabular}{l l c c r c}
\hline\hline
\\[-3mm]
Level\footnote{The fine structure of the D-series was not resolved} & Energy\footnote{From \citet{Risberg65}} & $\lambda_{\mathrm{exc}}$\footnote{Laser wavelength used to populate the upper state} & $\lambda_{\mathrm{obs}}$\footnote{Wavelength used to detect the fluorescence signal} & Lifetime\\
 & (cm$^{-1})$ & (\AA ) & (\AA ) & (ns)\\
\hline
\\[-3mm]
3s4s\hspace{0.7mm}$^{3}$S$_{1}$ & 41197.403 & 5183.65 & 5180 & 11.5(1.0)\\
3s5s\hspace{0.7mm}$^{3}$S$_{1}$ & 51872.526 & 3336.72 & 3340 & 29(3)\\
3s3d\hspace{0.7mm}$^{3}$D & 47957.0 & 3838.35 & 3840 & 5.9(4)\\
3s4d\hspace{0.7mm}$^{3}$D & 54192.3 & 3096.96 & 3100 & 17.6(1.2)\\
3s5d\hspace{0.7mm}$^{3}$D & 56968.2 & 2851.72 & 2850 & 33(3)\\
3p$^{2}$\hspace{0.7mm}$^{3}$P$_{0}$ & 57812.77 & 2781.49 & 2780 & 2.0(2)\\
3p$^{2}$\hspace{0.7mm}$^{3}$P$_{1}$ & 57833.40 & 2783.04 & 2780 & 2.0(2)\\
3p$^{2}$\hspace{0.7mm}$^{3}$P$_{2}$ & 57873.94 & 2779.90 & 2780 & 2.0(2)\\
\hline
\end{tabular}
\end{minipage}
\end{table}


\section{Branching fractions and wavelengths}

\subsection{Experimental method}

\begin{figure}
 \includegraphics[width=84mm]{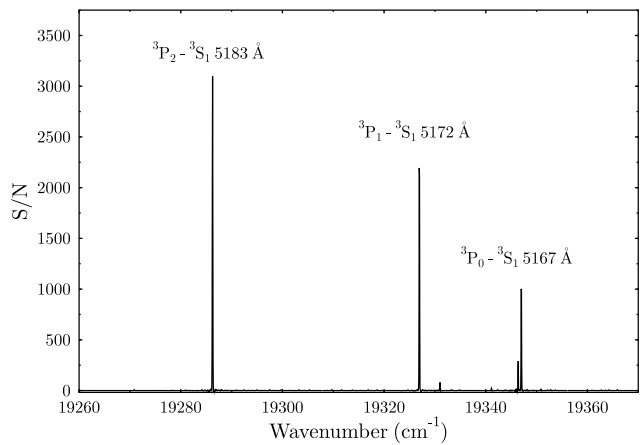}
 \caption{The observed 3p-4s \ion{Mg}{i} triplet. The line intensities in the figure are uncalibrated. The two weaker lines are from \ion{Fe}{i}.}
 \label{Tripletfig}
\end{figure}

The spectrum from \ion{Mg}{i} was produced in a water cooled hollow cathode discharge lamp (HCL). The cathode was made of pure iron with a small piece of magnesium placed in it. The cathode had an inner diameter of 7.0~mm and a length of 50~mm. Both neon and argon were tested to be used as the carrier gas in the HCL, but finally argon was chosen, since it has fewer strong lines in the measured wavenumber region. Choosing argon as the carrier gas also provided lines for wavenumber calibration, see Sect.~\ref{bfanalysis}. The light source was run with a current between 100 and 400~mA and a pressure between 0.6 and 0.7~torr. For all spectral acquisitions, within these light source parameters, the signal-to-noise ratios for the \ion{Mg}{i}~b triplet were good (SNR$>$100), see Fig.~\ref{Tripletfig}. When adjusting the parameters of the light source and the spectrometer, the HCL was run with a small piece of magnesium. The piece was later removed leaving only the layer of sputtered magnesium on the cathode walls. This was to decrease the density of Mg atoms and possible self-absorption of the stronger lines, see Sect.~\ref{selfabs}.

To correct the line intensities for the instrumental response an external continuous tungsten ribbon lamp was used. This has a known spectral intensity distribution for the wavelength interval 3000-7000 \AA . To ensure, to as high a degree as possible, that the alignment of the light from the HCL and the calibration lamp was the same, a T-setup with a folding mirror was used. This made it possible to switch between the two light sources without moving them. It is also crucial that the light sources are placed at the same distance from the aperture of the spectrometer so that the optical paths are equivalent.

The spectra were acquired with the Lund FT500 UV Fourier Transform (FT) spectrometer, which is optimized in the wavelength range of 2000-7000 \AA . Noise in the interferogram is transferred into the spectral region as white noise, with the same level throughout the spectrum. All spectral lines seen by the detector contribute to this noise level. It is therefore a disadvantage to record lines outside the region of interest. The detector, a Hamamatsu R928 photomultiplier, together with a hot-mirror filter were used to cut out light from outside the spectral region of interest. This also ensured that folding of the light from a neighbouring alias was eliminated, when using the continuous calibration lamp.

The three spectral lines from \ion{Mg}{i} 3p-4s are close in wavenumber, within $\sim$~60\,cm$^{-1}$. It was therefore only necessary to acquire spectra for one spectral region, 13000-25000\,cm$^{-1}$, with a resolution of 0.04\,cm$^{-1}$. This was sufficient to completely resolve the lines. Approximately twenty scans were co-added for each spectral acquisition to achieve good signal-to-noise ratios for the lines used in this study.

\subsection{Analysis}\label{bfanalysis}

The intensities of all lines from an upper level are used to determine the branching fractions of the transitions. The branching fraction is defined as the transition probability, $A_{ul}$, of a single line divided by the sum of all transition probabilities from the same upper level, $u$. Since the measured intensity of a line, $I_{ul}$, is proportional to the population of the upper level, $N_{u}$, the statistical weight of the upper level, $g_{u}$, and the transition probability, i.e.

\begin{equation}
I_{ul} \propto N_{u}g_{u}A_{ul}\,,
\label{Inteq}
\end{equation}

\noindent
the branching fraction can be expressed as

\begin{equation}
\left( BF \right)_{ul} = \frac{A_{ul}}{\sum_{l}A_{ul}} = \frac{I_{ul}}{\sum_{l}I_{ul}}\,.
\label{BFeq}
\end{equation}

\noindent
The measured intensities have however to be corrected for the instrument response, see Sect.~\ref{Intcalib}.

The analysis of the spectra was carried out using the FT spectrometry analysis computer program {\sc Xgremlin} \citep{Nave97b}, which is based on the {\sc Gremlin} code \citep{Brault89}. The spectral lines were fitted with Voigt profiles using a least-square procedure. Since Doppler broadening is the dominant broadening mechanism in the HCL, the fitted lines were expected to be close to pure Gaussians, which also was the case. The integrated area of the fitted profile was used as a measurement of the intensity of the spectral line.

Spectra observed by FT spectrometry have a linear wavenumber scale, whose accuracy derives from the control of the sampling of the interferogram by a single-mode helium-neon laser. The accuracy is however limited by the effects of using a finite-size aperture and by imperfect alignment of the light from the light source and the control laser \citep{Learner88}. To obtain a wavenumber scale, which is accurate to better than 1 part in $10^{7}$, a multiplicative correction is applied using a correction factor, $k_{\mathrm{eff}}$, such that
\begin{equation}
  \sigma_{\mathrm{corr}}=(1+k_{\mathrm{eff}})\sigma_{\mathrm{obs}}\,,
  \label{correq}
\end{equation}
where $\sigma_{\mathrm{corr}}$ is the corrected wavenumber and $\sigma_{\mathrm{obs}}$ is the observed, uncorrected, wavenumber.

The factor $k_{\mathrm{eff}}$ is accurately determined by measuring positions of one or several well-known internal wavenumber standard lines. In principle, it is possible to use only one calibration line \citep{Salit04}, but to reduce the uncertainty of the calibration, several calibration lines have been used for each recorded spectrum. Internal \ion{Ar}{ii} calibration lines, from the carrier gas in the HCL, where used for wavenumber calibration of the \ion{Mg}{i} spectra, where the wavenumbers of the \ion{Ar}{ii} calibration lines where taken from \citet{Whaling95}. The uncertainties of the measured wavenumbers are determined to be $\pm 0.002$~cm$^{-1}$, using the method described in \citet{aldenius06}.

\subsection{Intensity calibration}\label{Intcalib}

The instrument response of the FT spectrometer depends on the characteristics of the detector, the optical components and the optical path the light travels before entering the entrance aperture. Different methods to determine the instrument response are by using argon lines, with known branching ratios, generated in the light source itself or by using an external continuous calibration lamp. When using a hot-mirror filter, as has been done in this investigation, the response is however not as smooth and it is therefore crucial to measure the response continuously over the wavenumber region. We have used an external continuous tungsten ribbon lamp, which was placed at an equal distance from the FT spectrometer as the HCL. A folding mirror was used to switch between the tungsten lamp and the HCL. The tungsten lamp has been calibrated for spectral radiance by the Swedish National Testing and Research Institute to within 3\,\% (2$\sigma$) in the range 4000-8000 \AA\ . The recorded spectrum from the lamp is reproducible within 5\,\%.  In order to check that there had been no change in instrument response during the acquisition of the particular interferogram, calibration spectra were recorded immediately before and after each set of Mg interferograms. None of the acquisitions showed any significant differences between the two calibration spectra.

\begin{figure}
 \includegraphics[width=84mm]{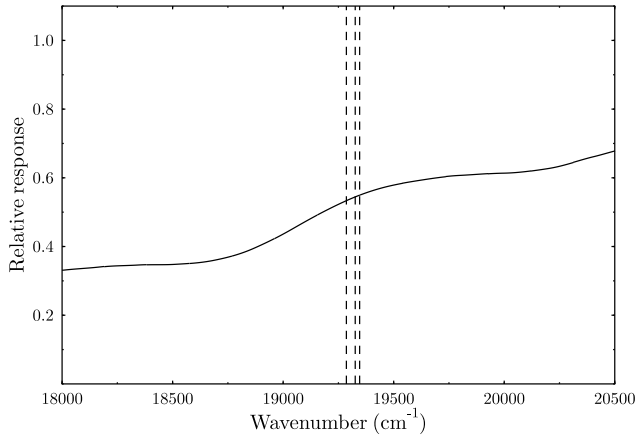}
 \caption{The instrumental response function (including optics, detectors and filter) for the wavenumber region of interest. The locations of the \ion{Mg}{i} triplet lines are marked in the figure with vertical dashed lines.}
 \label{Responsefig}
\end{figure}

The instrument response for the acquired spectra in the wavenumber region of interest is shown in Fig.~\ref{Responsefig}. The locations of the three spectral lines from 3p-4s transitions are marked with dashed lines. All three lines are within $\sim$~60\,cm$^{-1}$ and the response function varies slowly within this interval. The calibration could therefore be carried out with a very high accuracy, see Sect.~\ref{Uncertainties}.

\subsection{Self-absorption}\label{selfabs}

The three lower levels of the 3p-4s transitions have different ways to decay. The 3p~$^{3}$P$_{1}$ level has a decay channel with a spin-forbidden transition to the ground state and a lifetime of $\tau \sim 5$~ms \citep{Godone92}. The other two levels are metastable and thus have longer lifetimes, $\tau\sim 60$~min for 3p~$^{3}$P$_{2}$ and $\tau\sim 100$~min for $^{3}$P$_{0}$ \citep{Mishra01}. These two levels can therefore be highly populated and emitted photons might therefore be reabsorbed in the HCL. This will affect the observed intensity of the line by making it, and consequently the derived transition probability, too low.

Strong self-absorption not only lowers the intensity of a line, but might also distort the line profile. This is due to the fact that the absorption is stronger for the center of the line and the apparent line width will become larger than the normal Doppler width. The effect of increasing width can be seen as a bump in the residuals of the fitting of the line. If a line consists of unresolved contributions from several isotopes the profile may also become distorted since the more abundant isotopes will be more affected by self-absorption than the less abundant. Natural magnesium consists of one main isotope and two isotopes with smaller abundances ($^{24}$Mg:\,79\,\%, $^{25}$Mg:\,10\,\% and $^{26}$Mg:\,11\,\%). The transition shifts for the 3p-4s \ion{Mg}{i} lines have been measured, using a scanning Fabry-Perot interferometer, by \citet{Hallstadius77} reporting shifts of $-0.007$ and $-0.014$\,cm$^{-1}$, for $^{25}$Mg-$^{24}$Mg and $^{26}$Mg-$^{24}$Mg respectively. This is small compared to the Doppler width of about $0.1$\,cm$^{-1}$ in our observed lines. During the tests with different plasma densities in the HCL, no distortion or change in the line profiles was detected. To decrease the risk of strong self-absorption the magnesium piece was removed from the HCL leaving only the sputtered layer of Mg on the cathode walls. When fitting the Voigt profiles to the observed lines no visual self-absorption was detected. It was however necessary to further investigate this since a slight self-absorption is not always visible.

\begin{figure}
 \includegraphics[width=84mm]{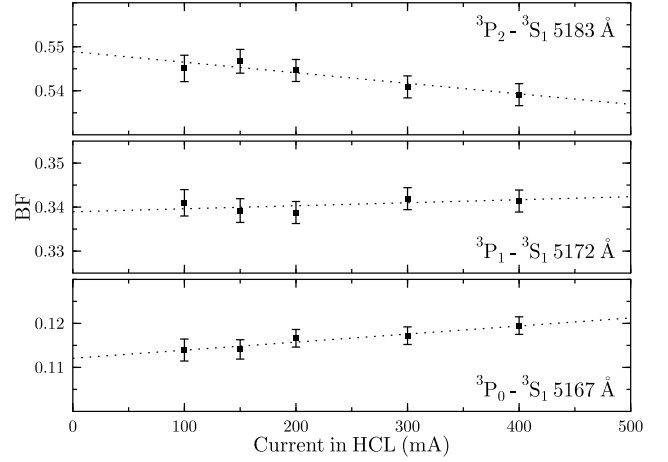}
 \caption{Test of optical thickness in the light source for the 3p~$^{3}$P-4s~$^{3}$S transitions in \ion{Mg}{i}. For an optically thin layer the branching fractions should not change with increasing current. The plot implies a slight self-absorption in the strongest line ($\lambda$5183 \AA ) at higher currents. The error bars represent the uncertainty in the branching fractions, not including the calibration uncertainties.}
 \label{Optfig}
\end{figure}

The population of the lower level depends on the plasma density and temperature. Different discharge currents in the HCL would produce different plasma densities and thus different amount of self-absorption. By measuring the line intensities at five different HCL currents, at 0.10, 0.15, 0.20, 0.30 and 0.40 A, the amount of self-absorption of the lines could be investigated. At zero current the lines are assumed to be free of self-absorption. The BFs for the three lines at different HCL currents are plotted in Fig.~\ref{Optfig}. The plot shows a slight self-absorption for the strongest, $\mathrm{\lambda}$5183~\AA , line, at higher currents. A straight line seems to represent the points well and by extrapolating to zero current the branching fractions were determined for an optically thin layer. Different ways to extrapolate and determine the optically thin branching fractions would change the BFs by less than 1~\% for the strongest line and by less than 5~\% for the weakest line.

\section{Oscillator strengths}

By combining the measured lifetime and branching fractions, the transition probabilities can be derived. Using the relation between the lifetime, $\tau$, and the transition probability,

\begin{equation}
\tau_{u} = \frac{1}{\sum_{l} A_{ul}}\,,
\label{Teq}
\end{equation}

\noindent
together with Eq.~\ref{BFeq} the expression for the transition probability becomes

\begin{equation}
A_{ul} = \frac{(BF)_{ul}}{\tau_{u}} \,.
\label{Aeq}
\end{equation}

\noindent
The absolute oscillator strength, $f$, can then be derived from the transition probability through

\begin{equation}
f = 1.499 \cdot 10^{-16} \frac{g_{u}}{g_{l}} \lambda^{2} A_{ul}\,,
\label{Osceq}
\end{equation}

\noindent
where $g_{u}$ and $g_{l}$ are the statistical weights for the upper and lower level, respectively, $\lambda$ is the wavelength in \AA , and $A_{ul}$ is in s$^{-1}$, see e.g. \citet{Thorne99,Huber86}.

\subsection{Uncertainties}\label{Uncertainties}

The uncertainty of the oscillator strength consists of contributions from the uncertainties of the branching fractions and the uncertainty of the radiative lifetime. The total uncertainties of the oscillator strengths are determined in the way described by \citet{Sikstrom02}. The uncertainty of the branching fraction of a transition depends on the uncertainties of the determination of the intensity of both that particular transition and of all other transitions from the same upper level. It also depends on the intensity calibration. In this case there are no missing lines or overlapping regions, since all three lines are measured in the same spectral region. The uncertainty of the correction of self-absorption contributes to the uncertainties of the branching fractions, due to the uncertainty in the extrapolation to zero current. This was estimated from the width, $\Delta y$, of the interval where the extrapolated curve may hit the ordinate axis in Fig.~\ref{Optfig}. Assuming a rectangular distribution this leads to a standard uncertainty of the correction being $\Delta y/3$.

The expression of the uncertainty, \emph{u}, of the absolute transition probability becomes
\begin{eqnarray}
\left( \frac{u(A_{ul})}{A_{ul}} \right)^{2} &=& \left( 1- (BF)_{ul}\right)^{2} \left( \frac{u(I_{ul}')}{I_{ul}'} \right)^{2} \nonumber \\
&+& \sum_{j\neq l}(BF)_{uj}^{2} \left[ \left( \frac{u(I_{uj}')}{I_{uj}'} \right) ^{2} + \left( \frac{u(c_{j})}{c_{j}} \right)^{2} \right] + \left( \frac{u(\tau)}{\tau} \right)^{2}\,,
\label{Unctoteq}
\end{eqnarray}
where $I'$ is the measured intensity and $c$ represents the intensity calibration. By also including the uncertainties from the correction of self-absorption the total uncertainty is derived. Due to the high accuracy of the branching fractions, the relative uncertainties of the oscillator strengths are low \mbox{($\sim$~1-3~\%),} whereas the absolute uncertainties depend on the uncertainty of the lifetime as well, adding up to $\sim$~9~\% for each of the three transitions.


\section{Results}\label{results}

In Table~\ref{Lifetable} we present the results of the measured lifetimes of 6 terms in the triplet series of \ion{Mg}{i} and comparison to previous measured lifetimes are presented in Table~\ref{Prevlifetable}, where the number in brackets gives the uncertainty in the final digit(s). The lifetimes of the doubly excited 3p$^{2}$\,$^{3}$P levels have not, to our knowledge, been measured before. The other levels in this investigation have been measured before, using different techniques. Our values are in general within the scatter of the previous values.  \citet{Kwiatkowski80} used a similar laser technique applied on an atomic beam oven producing the Mg ions, whereas \citet{Andersen72} used beam foil technique and \citet{Schaefer71} used delayed-coincidence technique.

\begin{table}
\begin{minipage}[t]{\columnwidth}
\caption{Comparison to previous experimental radiative lifetimes}
\label{Prevlifetable}
\centering
\renewcommand{\footnoterule}{}
\begin{tabular}{l c c c c}
\hline\hline
\\[-3mm]
 & \multicolumn{4}{c}{Experimental $\tau$ (ns)}\\ \cline{2-5}
 & This work & KTZ\footnote{\citet{Kwiatkowski80} using laser technique} & AMS\footnote{\citet{Andersen72} using beam foil technique} & S\footnote{\citet{Schaefer71} using delayed-coincidence technique} \\
\hline
\\[-3mm]
3s4s\hspace{0.7mm}$^{3}$S$_{1}$ & 11.5(1.0) & 9.7(6) & 10.1(0.8) & 14.8(0.7) \\
3s5s\hspace{0.7mm}$^{3}$S$_{1}$ & 29(3) & & & 25.6(2.1) \\
3s3d\hspace{0.7mm}$^{3}$D & 5.9(4) & 5.9(4) & 6.6(0.5) & 11.3(0.8) \\
3s4d\hspace{0.7mm}$^{3}$D & 17.6(1.2) & 15.6(9) & 13.5(1.0) & 18.4(0.7) \\
3s5d\hspace{0.7mm}$^{3}$D & 33(3) & 34.1(1.5) & & \\
3p$^{2}$\hspace{0.7mm}$^{3}$P$_{0}$ & 2.0(2) & & & \\
3p$^{2}$\hspace{0.7mm}$^{3}$P$_{1}$ & 2.0(2) & & & \\
3p$^{2}$\hspace{0.7mm}$^{3}$P$_{2}$ & 2.0(2) & & & \\
\hline
\end{tabular}
\end{minipage}
\end{table}

\begin{figure}
 \includegraphics[width=84mm]{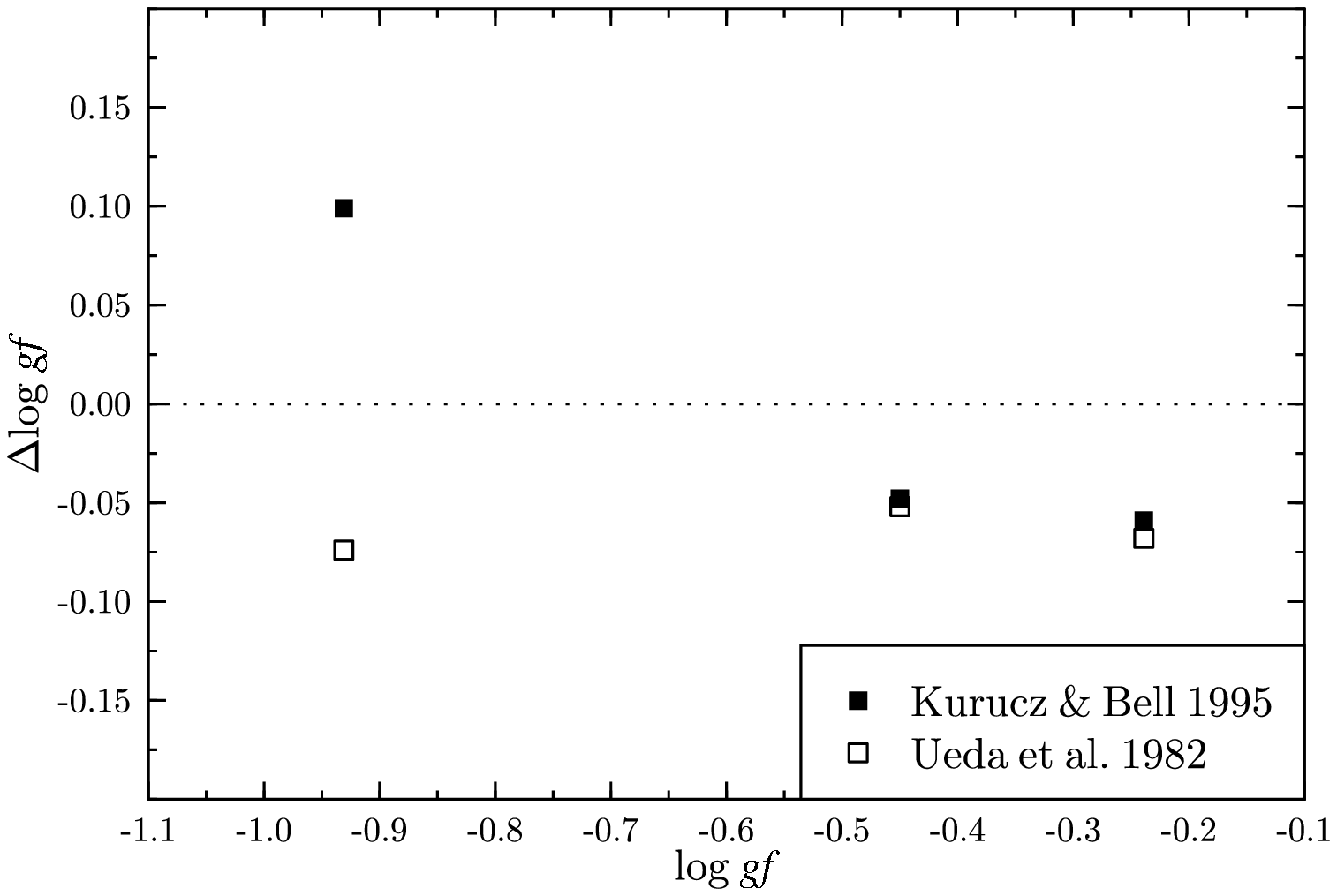}
 \caption{Comparison of the oscillator strengths from \citet{Kurucz95} and \citet{Ueda82} with the values derived in this work. $\Delta \mathrm{log}\,gf = \mathrm{log}\,gf_{\mathrm{\,This\,work}} - \mathrm{log}\,gf_{\mathrm{\,Prev.}}$}
 \label{Compfig}
\end{figure}

The results of the oscillator strengths, branching fractions, log~$gf$ values and wavelengths of the three 3p-4s transitions are presented in Table~\ref{Osctable} together with comparisons of previously published log~$gf$ values. The values of KB \citep{Kurucz95}, which are calculations by \citet{Kurucz75} and \citet{Anderson67}, are often used in stellar abundance determinations. The corresponding BF values show a larger scatter between the three lines than both our values and those of UKF \citep{Ueda82}, which are measured using the Hook method. The BF values of UKF are very similar to those presented in this work, however the absolute scale in UKF is based on the theoretical absolute $f$ values for \mbox{3p $^{3}$P} - \mbox{3d $^{3}$D} taken from \citet{Wiese69} and differs somewhat from ours. In Fig.~\ref{Compfig} the differences between the new $\mathrm{log}\,gf$ values and the values of KB and UKF are plotted.

\begin{table*}
\centering
\begin{minipage}[t]{17cm}
\caption{Experimental oscillator strengths of the \ion{Mg}{i} 3p-4s triplet. $\lambda_{\mathrm{air}}$ is derived using the \citet{Edlen66} dispersion formula.}
\label{Osctable}
\renewcommand{\footnoterule}{}
\centering
\begin{tabular}{l c c c c c c c c c}
\hline\hline
Transition & $\sigma$ & $\lambda_{\mathrm{air}}$ & $BF$ & Uncert. & $gf$ & Uncert. & \multicolumn{3}{c}{$\log gf$} \\ \cline{8-10}
& (cm$^{-1}$) & (\AA ) & & (\% in $BF$) & & (\% in $gf$) & This work & KB\footnote{Compiled in Kurucz CD-ROM No. 23 \citep{Kurucz95}} & UKF\footnote{Measurements by \citet{Ueda82} using the Hook method} \\
\hline
\\[-3mm]
3s3p $^{3}$P$_{2}$ - 3s4s $^{3}$S$_{1}$ & 19286.224 & 5183.6046 & 0.549 & 1 & 0.577 & 9 & -0.239 & -0.180\footnote{Calculations by \citet{Anderson67}} & -0.171 \\
3s3p $^{3}$P$_{1}$ - 3s4s $^{3}$S$_{1}$ & 19326.938 & 5172.6847 & 0.339 & 2 & 0.355 & 9 & -0.450 & -0.402$^c$ & -0.399 \\
3s3p $^{3}$P$_{0}$ - 3s4s $^{3}$S$_{1}$ & 19346.994 & 5167.3222 & 0.112 & 3 & 0.117 & 9 & -0.931 & -1.030\footnote{Calculations by \citet{Kurucz75}} & -0.857 \\
\hline
\end{tabular}
\end{minipage}
\end{table*}


\section{Astrophysical applications}\label{astrophyssec}

The \ion{Mg}{i} b triplet is very prominent in the optical spectra of solar-type stars, being amongst the strongest metal absorption lines apart from the \ion{Ca}{ii} H and K doublet. Even at moderate spectral resolution ($\sim$ 1~\AA) and S/N ($\sim$ 10), they are strong enough to be visible in many of the low-metallicity giants in the Hamburg-ESO survey. In contrast to the H and K lines which have traditionally been used to identify very metal-poor stars spectroscopically, the b lines lie in a spectral region closer to the flux peak of solar-type and cooler stars, and where CCDs generally have higher quantum efficiency. These factors, combined with their visibility even in metal-poor stars in the Hamburg-ESO survey, makes them potentially valuable alternatives to the violet H and K lines in the identification of low-metallicity stars. Indeed, the spectral region selected by \citet{Carney87} in their high resolving-power, low-S/N study of metal-poor stars included the b lines.

\begin{figure}
 \includegraphics[angle=270,width=84mm]{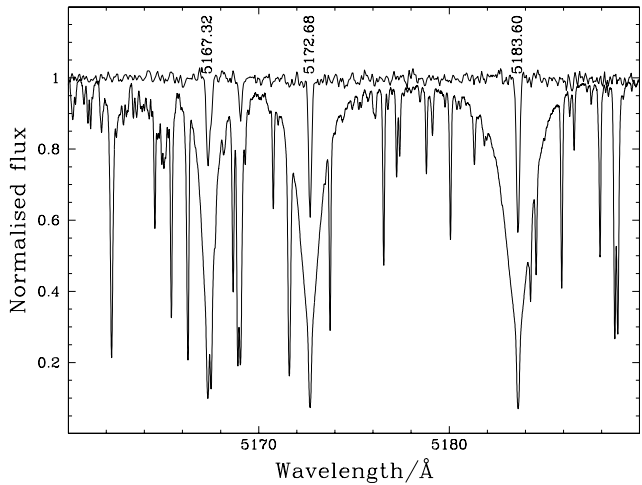}
 \caption{Spectrum of the Sun and the metal-deficient star CD~$-$24$^\circ$17504 in the region of the \ion{Mg}{i} b lines. The signal-to-noise ratio in the latter star is approximately 100, giving noticeable noise in the continuum, whereas the solar spectrum is almost noise-free. The 5167~\AA\ member of the triplet is blended with an iron line in both stars, but the two longer wavelength transitions are free of significant blends in the metal-deficient star.}
 \label{Astrofig}
\end{figure}

Fig.~\ref{Astrofig} shows the b line spectral region for the Sun \citep{Beckers76} and the metal-deficient dwarf or subgiant CD~$-$24$^\circ$17504, with effective temperatures of 5770~K and 6070~K respectively. The latter star has a magnesium abundance only 6.8$\times$10$^{-4}$ times the solar value, but the b lines nevertheless can be readily measured \citep{Arnone05}.

Two recent uses of the b lines in high-resolution studies are those of \citet{Arnone05} and \citet{Aoki06}. The former investigates the uniformity of the Mg/Fe ratio in 23 low-metallicity dwarfs, while the latter analysis is of the currently most iron-poor star known,  HE~1327-2326. The Arnone et al. study was principally concerned with star-to-star differences in the Mg/Fe ratio, and hence to first order it was immune to errors in the b-line $gf$ values since similar Mg lines are used in each star. However, when these data are compared with measurements of the Mg/Fe ratio in other studies in which different Mg lines are observed (such as the three similarly strong violet lines at 3829-3838~\AA), or are compared with theoretical calculations of supernova yields, then it is important that the $gf$ values are known reliably. Because of the paucity of metals in HE~1327-2326, the Mg abundance derived for this star hinges strongly on the b-line $gf$ values.

As noted in the introduction to this paper, previously only one \ion{Mg}{i} transition had a laboratory determined $gf$ value, with inverse solar analyses and theoretical calculations accounting for the remainder. The $gf$ values derived for the b lines in the present work (Table~\ref{Osctable}) from measurements of level lifetimes and branching fractions now put the b line data on a very secure footing, with uncertainties of only 9~\% , i.e. 0.04 in log $gf$.
Fortunately, the values measured are very similar to those used by \citet{Arnone05}, viz. $-$0.17 and $-$0.39 for 5183 and 5172~\AA\ respectively, and by \citet{Aoki06}, viz. $-$0.180, $-$0.402 and $-$1.030. 
Our measurements give values lower by 0.06-0.07 for 5183~\AA, lower by 0.05-0.06 for 5172~\AA, and higher by 0.10 for 5167~\AA. Switching the cited analyses to our $gf$ values would increase the [Mg/Fe] values typically by 0.04 in \citet{Arnone05} (since generally the two longer-wavelength b lines and one other Mg line were used), and would have no significant net effect ($\sim$ 0.01) on [Mg/Fe] in \citet{Aoki06}. Such revisions are within the small (0.04 dex) uncertainty in the newly measured $gf$ values, and are comparable with or smaller than other typical uncertainties in the spectral analyses. The small size of these corrections, besides being a relief, also provides valuable verification of the inverse solar analysis by \cite{Fuhrmann95} and the theoretical calculations by \citet{Chang90} upon which \citet{Arnone05} relied for $gf$ values. 


\section{Conclusions}\label{conclusionsect}

Measurements of stellar abundances in stars rely on the log\,$gf$ values of the spectral lines.
Prompted by the need for accurate log\,$gf$ values in abundance analyses of metal-poor stars, oscillator strengths for the optical \ion{Mg}{i} 3p-4s triplet have been determined by combining branching fractions, measured with Fourier Transform spectrometry, with radiative lifetimes, measured with laser induced fluorescence. The uncertainties in the $f$ values are determined to be $9$~\% for the three lines. This corresponds to an uncertainty of $\pm0.04$~dex in the log\,$gf$ values. This will lead to greater confidence in the abundance analyses of magnesium in metal-poor stars. In total, radiative lifetimes have been measured for the 3s4s\,$^3$S, 3s5s\,$^3$S, 3s3d\,$^3$D, 3s4d\,$^3$D, 3s5d\,$^3$D and 3p$^2$\,$^3$P terms in \ion{Mg}{i}. The log~$gf$ values can be found in Table~\ref{Osctable} and the radiative lifetimes can be found in Table~\ref{Lifetable}. The values have been compared to previous measurements and calculations, when possible. Laboratory wavenumbers and wavelengths for the 3p-4s lines are also reported, with an uncertainty of $\pm0.002$\,cm$^{-1}$, corresponding to $\pm0.5$\,m\AA .

\begin{acknowledgements}
We thank Dr~H.~Xu for his assistance in the laboratory during the lifetime measurements and Dr~H.~Nilsson for discussions regarding the branching fraction measurements.
S.G.R would also like to thank Dr~J.~Pickering for discussions relating to this work and Dr~S.~Feltzing for funding an earlier visit to Lund which initiated this collaboration.
The work was supported by the EU-TMR access to Large-Scale Facility Programme (contract RII3-CT-2003-506350).
\end{acknowledgements}


\bibliographystyle{aa}


\end{document}